\documentclass[12pt]{article}

\def\be{\begin{equation}}
\def\ee{\end{equation}}
\def\bea{\begin{eqnarray}}
\def\eea{\end{eqnarray}}

\def\e{{\rm e}}

\def\d{{\rm d}}

\usepackage{graphicx}
\usepackage{bm}
\usepackage{amsmath}
\usepackage{amssymb}

\begin{document}

\begin{flushright}
astro-ph/0607257
\end{flushright}

\pagestyle{plain}

\begin{center}
\vspace{2.5cm} {\Large {\bf Noncommutative Black-Body Radiation: \\ \vspace{0.5cm} Implications On Cosmic Microwave Background}}

\vspace{1cm}

Amir H. Fatollahi$^{~(1}$  \hspace{3mm} and \hspace{3mm} Maryam Hajirahimi$^{~(2}$

\vspace{.5cm}

{\it 1) Mathematical Physics Group, Department of Physics, Alzahra University, \\ P. O. Box 19938, Tehran 91167, Iran}

\vspace{.3cm}

{\it 2) Institute for Advanced Studies in Basic Sciences (IASBS),\\
P. O. Box 45195, Zanjan 1159, Iran}

\vspace{.3cm}

\texttt{fatho@mail.cern.ch\\
rahimi@iasbs.ac.ir}

\vskip .5 cm
\end{center}

\begin{abstract}
Including loop corrections, black-body radiation in noncommutative space is anisotropic.
A direct implication of possible space noncommutativity on the Cosmic Microwave Background map is argued.
\end{abstract}

\vspace{2cm}

\newpage

There have been arguments supporting the idea that the ordinary picture of spacetime breaks down when
is probed with sufficiently large momenta and energies. In particular, in an ultra-large momentum
transfer experiment a black-hole may be formed, and as long as it lives before its rapid evaporation,
an observer experiences limits on information transfer from the volume element comparable in size with the horizon
\cite{doplicher}. These kinds of reasoning may lead one to believe in some kinds of space-space and
space-time uncertainty relations \cite{doplicher}. As uncertainty relations usually
point to noncommutative objects, it is reasonable to consider various versions of noncommutative spacetime
theories, among them theories defined on spacetime whose coordinates satisfy the canonical relation
\bea\label{algebra}
[\widehat{x}^{\,\mu},\widehat{x}^{\,\nu}]=i\lambda^{\mu\nu},
\eea
in which $\lambda^{\mu\nu}$ is an antisymmetric constant tensor.
Via recent developments in understanding the dynamics of D-branes of string theory,
there has been a renewed interest for studying field theories on spacetimes
whose coordinates satisfy the above algebra.
In particular, the longitudinal directions of D-branes in the presence of constant B-field
background appear to be noncommutative, as are seen by the ends of open
strings \cite{9908142}.

The phenomenological implications of possible noncommutative coordinates have been considered in
a very large number of works. Among many others, here we can give just a
brief list of works, and specially those concerning the phenomenological implications of noncommutative QED.
The effect of noncommutativity of space is studied for possible modifications
that may appear in high energy scattering amplitudes of particles \cite{phen-nc1}, in energy levels of
light atoms \cite{phen-nc2, phen-nc3}, and anomalous magnetic moment of electron \cite{jab3}.
The ultra-high energy scattering of massless photons of noncommutative U(1) theory is considered
in \cite{mahajan} and the tiny change in the total amplitude is obtained as a function of the total energy.
Some other interesting features of noncommutative ED and QED are discussed in \cite{morencqed,jackiw}.

In present work we address the radiation we expect from a black-body in noncommutative space.
As we deal with a black-body radiation problem, the natural framework is finite temperature field theory.
Noncommutative QED, though renormalizable, shares features suggesting that the present formulation of
theory possibly has to be modified to be considered as a true theory. In particular, the IR limits of
physical quantities are irregular once they are compared with their counterparts in theories defined
in ordinary space. In spite of these mentioned difficulties, one might be hopeful that the results
obtained based on the present formulation can still offer a sense
for what we should expect as an indication of noncommutativity, if any after all.

It is understood that field theories on noncommutative spacetime are defined by actions that
are essentially the same as in ordinary spacetime, with the exception that the products
between fields are replaced by $\star$-product, defined for two functions $f$ and $g$  \cite{reviewnc}
\bea
(f\star g)(x)=\exp\big(\frac{i\lambda^{\mu\nu}}{2}\partial_{x_\mu}\partial_{y_\nu}\big)f(x)g(y)\mid_{y=x}
\eea
It can be seen that the $\star$-product is associative, {\it i.e.}, $f\star g\star h=(f\star g)\star h= f\star (g\star h)$,
and so it is not important which two should be multiplied firstly.
Though $\star$-product itself is not commutative ({\it i.e.}, $f \star g \neq g \star f$), we have
$\int f\star g=\int g\star f=\int fg$, saying in integrands always one of the stars can be removed.

The pure gauge field sector of noncommutative U(1) theory is defined by the action
\bea
S_{\rm gauge-field}=-\frac{1}{4}\int {\rm d}^{4}x \;F_{\mu\nu}\star F^{\mu\nu}=-\frac{1}{4}\int {\rm d}^{4}x \;F_{\mu\nu}F^{\mu\nu}
\eea
with $F_{\mu\nu}=\partial_{\mu}A_{\nu}-\partial_{\nu}A_{\mu}-ie[A_{\mu},A_{\nu}]_{\star}$,
by definition $[f,g]_{\star}=f\star g-g\star f$.
The action above is invariant under local gauge symmetry transformations
\bea\label{trans}
A'_{\mu}= U\star A_{\mu}\star U^{-1}+\frac{i}{e}U\star \partial _{\mu}U^{-1}
\eea
in which $U=U(x)$ is the $\star$-phase, defined by a function $\rho(x)$ via the $\star$-exponential:
\bea\label{starphase}
U(x)=\exp_{\star}(i\rho)=1+i\rho-\frac{1}{2}\rho\star\rho+\cdots,
\eea
with $U^{-1}=\exp_{\star}(-i\rho)$, and $U\star U^{-1}=U^{-1}\star U=1$. Under above transformation, the field strength transforms as
$F_{\mu\nu}\to F^{\prime}_{\mu\nu}=U\star F_{\mu\nu}\star U^{-1}$.
We mention that the transformations of gauge field as well as the field strength
look like those of non-Abelian gauge theories. Besides we see that the
action contains terms which are responsible for interaction
between the gauge particles. We see how the noncommutativity of coordinates induces properties on fields and
their transformations, as if they were belong to a non-Abelian theory;
the subject that how the characters of coordinates and fields may be related to each other
is discussed in \cite{fath}.

The other interesting feature of field theories defined by $\star$-product is that these theories exhibit some
aspects very reminiscent of string theory. In particular, in these theories the quanta of fields interact as
extended objects, namely electric dipoles \cite{jab5}. Also in these kinds of field theories one recognizes
much more distinct role and behavior than ordinary theories for planar and non-planar Feynman diagrams \cite{nc-planarity}.

As we deal with a black-body radiation problem, the natural framework is finite temperature field theory
\cite{fin-temp-book}. Finite temperature noncommutative field theory has been the subject of research
works \cite{9912140,tnc-1,tnc-2}. As mentioned, noncommutative U(1) gauge theory is involved by
self-interaction of photons, and so beyond the free theory one finds deviations from the expression by ordinary
U(1) theory for black-body radiation. The Feynman rules of noncommutative U(1) theory are known \cite{tnc-2}.
Here we consider noncommutativity only for spatial directions, assuming
$\lambda^{0i}=0$. By this one can use the expressions already derived for non-Abelian gauge theory \cite{kapusta}, except that
here the vertex-functions are momentum dependent. The expression for free-energy in unit volume at
temperature $T$ at two-loop order is given by ($\hbar=c=1$, $\beta=T^{-1}$) \cite{9912140}:
\bea
{\cal F}(T)= {\cal F}_{\rm isotropic}+4e^2\int\frac{\d ^3k}{(2\pi)^3}\int
\frac{\d ^3k'}{(2\pi)^3}\;\frac{\sin^2({\bf k}\ltimes{\bf k'})}{\omega(\e^{\beta \omega}-1)\omega'(\e^{\beta \omega'}-1)}
\eea
in which ${\cal F}_{\rm isotropic}$ represents the part that does not depend on the noncommutativity parameter.
As we shall see the other part results in an energy flow that depends on direction.
In above $\omega=|{\bf k}|$ and $\omega'=|{\bf k'}|$, and ${\bf k}\ltimes{\bf k'}=\frac{1}{2}\lambda^{ij}k_ik'_j$.
We mention, as pointed earlier, the expression is convergent both in IR ($\omega,\omega'\to 0$) and
UV ($\omega,\omega'\to \infty$) limits. For the more important IR limit, the reason comes
back to the fact that noncommutativity scale effectively cuts off interactions at large distances \cite{9912140}.
Using the relation $U(T)={\cal F}-T\partial_{\,T} {\cal F}$, we have for the energy-density $U(T)$
\bea
U(T)&=&U_{\rm isotropic}
+4e^2\int\frac{\d ^3k}{(2\pi)^3}\int\frac{\d ^3k'}{(2\pi)^3}\frac{\sin^2({\bf k}\ltimes{\bf k'})}{\omega\omega'(\e^{\beta\omega}-1)(\e^{\beta\omega'}-1)}
\nonumber\\
&&\!\!\!\!\!\!\!\!-4\beta e^2\int\frac{\d ^3k}{(2\pi)^3}\int\frac{\d ^3k'}{(2\pi)^3}
\frac{\sin^2({\bf k}\ltimes{\bf k'})}{(\e^{\beta\omega}-1)(\e^{\beta\omega'}-1)}
\Big(\frac{\e^{\beta\omega}}{\omega'(\e^{\beta\omega}-1)}
+\frac{\e^{\beta\omega'}}{\omega(\e^{\beta\omega'}-1)}\Big)
\eea
One may define the vector $\bm{\lambda}$ by its components $\lambda_k=\frac{1}{4}\epsilon_{ijk}\lambda^{ij}$. By taking
$\bm{\lambda}=\lambda\,{\bf z}$, one finds ${\bf k}\ltimes{\bf k'}=\bm{\lambda}\cdot({\bf k}\times{\bf k'})=\lambda({\bf k}\times{\bf k'})_z$,
and so
\bea
{\bf k}\ltimes{\bf k'}=\lambda\,\omega\omega'\sin\theta\sin\theta'\sin(\phi'-\phi)
\eea
in which ${\bf k}$ and ${\bf k'}$ are given as ${\bf k}=(\omega,\theta,\phi)$ and ${\bf k'}=(\omega',\theta',\phi')$
in spherical coordinates. By the Taylor expansion of $\sin^2\alpha$ \cite{gr-rez}, and taking $\alpha={\bf k}\ltimes{\bf{k'}}$,
via (8) one has
\bea
\sin^2({\bf k}\ltimes{\bf{k'}})=\sum^{\infty}_{m =1} a_m \sin^{2m}\!\theta'\sin^{2m }(\phi'-\phi)
\eea
in which $a_m =\displaystyle{\frac{(-1)^{\rm m+1}2^{2m -1}}{(2m )!}(\lambda\omega\omega'\sin\theta)^{2m}}$. So we find
\bea
\int \d\Omega'\sin^2({\bf k}\ltimes{\bf k'})
=\sum^{\infty}_{m =1}\frac{\pi(-1)^{\rm m+1}2^{2m +1}}{(2m +1)!}(\lambda\omega\omega'\sin\theta)^{2m }
\eea
For $U(T,\Omega)$ as the energy-density received from the solid-angle $\d \Omega$ we find
\bea
U(T,\Omega)\,\d \Omega=\bigg[\frac{\sigma_0}{4\pi} T^4
+\frac{4e^2}{(2\pi)^6}\sum^{\infty}_{m =1}\frac{\pi(-1)^{\rm m+1}2^{2m +1}}{(2m +1)!} (\lambda^{2m })
T^{4m+4}I_m \sin^{2m}\!\theta\bigg] \;\d \Omega
\eea
in which $\sigma_0=\frac{\pi^2}{15}$ (Stefan's constant=$\frac{\pi^2}{60}$), and
\bea
I_m =\bigg(\int_0^\infty\frac{s^{2m +1}\d s}{(\e^{s}-1)}\bigg)^2
-2\int_0^\infty\frac{s^{2m +2}\e^{s}\d s}{(\e^{s}-1)^2}\int_0^\infty\frac{s'^{2m +1}\d s'}{(\e^{s'}-1)}.
\eea
One finds $I_m =-(4m +3)\zeta^2(2m +2)\big((2m +1)!\big)^2$ by the following relations
\bea
\int_0^\infty\frac{s^{2m +1}\d s}{(\e^{s}-1)}=\zeta(2m +2)(2m +1)!,
\\
\int_0^\infty\frac{s^{2m +2}\e^{s}\d s}{(\e^{s}-1)^2}=\zeta(2m +2)(2m +2)!,
\eea
with $\zeta(t)$ as the Riemann zeta-function. Finally we have
\bea
U(\Omega, T)=\frac{\sigma_0}{4\pi} T^4-\frac{4\pi e^2}{(2\pi)^6}T^4\sum^{\infty}_{m =1}(\lambda T^2)^{2m} J_m\sin^{2m}\!\theta
\eea
in which $J_m=(-1)^{m +1}2^{2m +1}(4m +3)\zeta^2(2m +2)(2m +1)!$. In leading order one has
\bea\label{fin-expr}
U(\Omega, T)=\frac{\sigma_0}{4\pi} T^4-\frac{7\pi^4}{675}\alpha \,T^4(\lambda T^2)^2 \sin^2\!\theta + O\big((\lambda T^2)^4\big)
\eea
in which $\alpha=\frac{e^2}{4\pi}\simeq \frac{1}{137}$.

We mention as much as one comes out from the noncommutative direction, the radiation is decreasing, giving the minimum
for $xy$-plane, $\theta=\frac{\pi}{2}$. This is simply due to the fact that the coupling of photons is related to their momenta.
In particular the photons moving in the plane perpendicular to noncommutative direction feel the strongest coupling with respect to others,
yielding a decrease in outgoing radiation.

The final comment is about the possible contribution of fermionic degrees of freedom to anisotropy.
In fact one can check easily that, as the vertex-function for coupling of fermions to gauge fields depends on $\lambda$
only through a phase factor \cite{tnc-2}, the expression coming from fremionic degrees of freedom, due to cancellation of two phases,
is isotropic.

Although it is hard to imagine that the implications of noncommutativity can be detected in
a laboratory black-body radiation, one may look for an indication of noncommutativity
in the signals we are getting from the extremely hot seconds of early universe.
In fact, the energy scale that one expects for relevance of noncommutative effects is as much as
high and this suggests maybe it has been available for particles only in the early universe.
So an excellent way to test the phenomenon related to noncommutativity of spacetime would be the study of
what are left for us as early universe's heir, the most important among them the Cosmic Microwave Background
(CMB) radiation. The reason is, CMB map is just a tableau of events which happened at the first epochs of
universe, at the decoupling era or much earlier, when the energies were sufficiently high to make relevant
possible spacetime noncommutativity. In \cite{tx-uncer} the consequences of space-time uncertainty relations of the form
$\Delta t \Delta x\geq l_s^2$ are studied in the context of inflation theory, and possible applications of these relations
in better understanding of present CMB data are discussed. Also there have been efforts to formulate and study the
noncommutative versions of inflation theory \cite{nc-cosmos}. In \cite{fuzzy-sphere} by taking the blowing sphere
that eventually plays the role of the so-called last-scattering surface as a fuzzy sphere
some kinds of explanation is presented for the relatively low angular power spectrum $C_l$ in small $l$ region ($l\simeq 6)$.
Recently in \cite{nc-field} it was studied that how a theory with noncommutative electromagnetic fields - that is considering the fields, rather
than the coordinates, noncommutative - may change the pattern we expect to see in polarized CMB data.

As CMB map is in fact nothing more than a black-body radiation pattern which is slightly perturbed by
fluctuations, instead of dealing with the implications of noncommutativity on
different cosmological models, here we can directly address what one should expect to see
in CMB map if in early universe the coordinates had satisfied the algebra (\ref{algebra}).
According to the expression we obtained, space noncommutativity in early universe modifies the pattern we expect to see in the CMB map sky. Replacing $\sin^2\!\theta$
by a combination of $P_0(\cos\theta)$ and $P_2(\cos\theta)$ as zeroth and second Legendre polynomials respectively,
in leading order the noncommutative effects modify the monopole and quadrupole moments of angular power spectrum. The temperature
$T_0$ associated to the black-body, defined by $\int U(\Omega, T)\d \Omega = \sigma_0 T_0^4$, is modified by the monopole term.
The term proportional to $P_2(\cos\theta)$ results in an anisotropy in the measured power as well as the temperature $T(\Omega)$
that one may associate to the radiation received from the solid-angle $\d \Omega$. The temperature $T(\Omega)$ is defined by
$T(\Omega)=\big(\frac{4\pi}{\sigma_0} U(\Omega, T)\big)^{1/4}$.
Much effort is currently being devoted to examining the CMB temperature anisotropies measured with the Wilkinson Microwave
Anisotropy Probe (WMAP) \cite{wmap}. It would be extremely important if the present and forthcoming data indicated any
significant evidence for canonical noncommutativity in the early universe, a thing which could count anisotropic radiation
among its direct implications.

\vspace{0.2cm}
{\bf Acknowledgement:}
A. H. F. is grateful to M. Khorrami, and specially to A. Hajian for very helpful discussions.


\end{document}